\documentclass[5p,times]{elsarticle}

\usepackage{siunitx}
\usepackage{amsmath}
\usepackage{wasysym}
\usepackage{graphicx}
\usepackage{xcolor}
\usepackage{xfrac}
\usepackage{tablefootnote}
\usepackage{natbib}

\newcommand{\penD}{\lambda_{\rm{L}}}
\newcommand{\muEff}{\mu_{\rm{r,eff}}}
\newcommand{\epsEff}{\varepsilon_{\rm{r,eff}}}
\newcommand{\microEV}[1]{\SI{#1 }{\micro\eV}}
\newcommand{\microO}[1]{\SI{#1}{\micro\ohm}}
\newcommand{\micron}[1]{\SI{#1}{\micro\metre}}
\newcommand{\microH}[1]{\SI{#1}{\micro\henry}}
\newcommand{\cm}{cm$^{-1}$}

\usepackage{lineno,hyperref}
\modulolinenumbers[5]

\journal{Journal of \LaTeX\ Templates}

\begin{document}

\begin{frontmatter}

\title{Superconducting boron doped nanocrystalline diamond microwave coplanar resonator}

\author[CU-PHYSX]{Jerome A. Cuenca\corref{mycorrespondingauthor}}\ead{cuencaj@cardiff.ac.uk}
\author[CU-PHYSX]{Thomas Brien}
\author[CU-PHYSX]{Soumen Mandal}
\author[CU-PHYSX]{Scott Manifold}
\author[CU-PHYSX]{Simon Doyle}
\author[CU-ENGIN]{Adrian Porch}
\author[CU-PHYSX]{Georgina M. Klemencic}
\author[CU-PHYSX]{Oliver A. Williams}
\address[CU-PHYSX]{School of Physics and Astronomy, Cardiff University, Cardiff, Wales, CF24 3AA, UK}
\address[CU-ENGIN]{School of Engineering, Cardiff University, Cardiff, Wales, CF24 3AA, UK}
\cortext[mycorrespondingauthor]{Corresponding author}

\begin{abstract}
A superconducting boron doped nanocrystalline diamond (B-NCD) coplanar waveguide resonator (CPR) is presented for kinetic inductance ($L_k$) and penetration depth ($\penD$) measurements at microwave frequencies of 0.4 to 1.2 GHz and at temperatures below 3 K. Using a simplified effective medium CPR approach, this work demonstrates that thin granular B-NCD films ($t\approx $ \SI{500}{\nano\metre}) on Si have a large penetration depth ($\penD\approx 4.3$ to \micron{4.4}), and therefore an associated high kinetic inductance ($L_{k,\square} \approx  $ 670 to 690 pH/$\square$). These values are much larger than those typically obtained for films on single crystal diamond which is likely due to the significant granularity of the nanocrystalline films. Based on the measured Q factors of the structure, the calculated surface resistance in this frequency range is found to be as low as $\approx$ 2 to \microO{4} at $T<2$ K, demonstrating the potential for granular B-NCD for high quality factor superconducting microwave resonators and highly sensitive kinetic inductance detectors.

\end{abstract}


\end{frontmatter}

\section{Introduction}
Boron doped diamond (BDD) is electrically conducting at concentrations in excess of $10^{20}$ cm$^{-3}$\cite{Gajewski2009,Cobb2018} and superconducts at low temperatures in excess of $10^{21}$ cm$^{-3}$, with an observed critical onset temperature ($T_c$) ranging between 3 to 5 K at zero field and a high type-II upper critical field ($H_c$) of up to 8 T\cite{Ekimov2004,Winzer2005,Takano2004,Bustarret2004,Mandal2011,Kageura2019,Klemencic2017}. BDD is typically grown using chemical vapour deposition (CVD) either on single crystal diamond substrates (B-SCD) or on Si to produce heavily granular nanocrystalline diamond films (B-NCD). These films can be made into superconducting quantum interference devices (SQUIDs) as demonstrated by Kageura et al. on B-SCD\cite{Kageura2019} and in earlier reports by Mandal et al. on B-NCD\cite{Mandal2010b,Mandal2010a,Mandal2011}. In order to understand the effective areas of BDD SQUIDs and also utilise BDD for other superconducting applications, such as microwave devices, further understanding of the magnetic penetration depth ($\penD$), is needed. For B-SCD, moderate penetration depths have been observed ($\penD \approx$ 0.2 to \micron{1})\cite{Winzer2005,Ortolani2006}. However, for granular B-NCD, a recent study by Oripov et al. has demonstrated very large penetration depths ($\penD\approx2$ to \micron{4}) in the microwave frequency range, thereby enabling highly sensitive microwave kinetic inductance ($L_k$) detectors\cite{Oripov2021}. The origin for the large penetration depths in B-NCD is most likely related to the granularity of the superconductor which is observed for other materials; for example, non-granular aluminium films exhibit very small penetration depths ($\penD\approx$ 10 to \SI{30}{\nano\metre}\cite{McLean1962}), however granular aluminium films have much larger values ($\penD\approx$ 0.3 to \micron{1.2}\cite{Cohen1968a}). 

A method for investigating the penetration depth and kinetic inductance of materials is by using the coplanar waveguide resonator (CPR) method\cite{Watanabe1994,Porch2005}. This approach involves patterning the superconductor into a planar microwave structure with a known resonant frequency and attributing differences to the designed frequency to the kinetic inductance, and therefore the penetration depth. Since high quality factor resonators can be fabricated, this approach offers a highly sensitivity measurement. Additionally, the CPR method is very convenient for CVD films since a CPR only requires one side of a substrate to patterned without the need for any additional ground planes on the underside. The CPR method is less-widely reported for superconducting BDD, despite being a well-known approach for understanding $L_k$ in a wide variety of materials including aluminium\cite{Doyle2008,Noguchi2018,Zhang2019a}, niobium nitride\cite{Watanabe1994} and yttrium barium copper oxide\cite{How1992b,Porch1995,Yoshida1995}.

In this work the seemingly high $L_k$ and $\penD$ of granular B-NCD is investigated using the microwave CPR method. Section \ref{sec:theory} details a simplified analytical model for the approximation of $L_k$ in a CPR using an effective medium. Section \ref{sec:model} briefly demonstrates a finite element model (FEM) to identify the resonant modes and implement the effective medium approach. Section \ref{sec:method} details the device fabrication and characterisation process. Finally, Sections \ref{sec:exp} and \ref{sec:disc} details the experimental results, including Raman spectroscopy, scanning electron microscopy (SEM) and microwave power transmission spectra.


\section{Theory}\label{sec:theory}
\begin{figure}[t]
	\centering
	\includegraphics[trim={8.5cm 5.5cm 8.5cm 0},clip, width=0.4\textwidth]{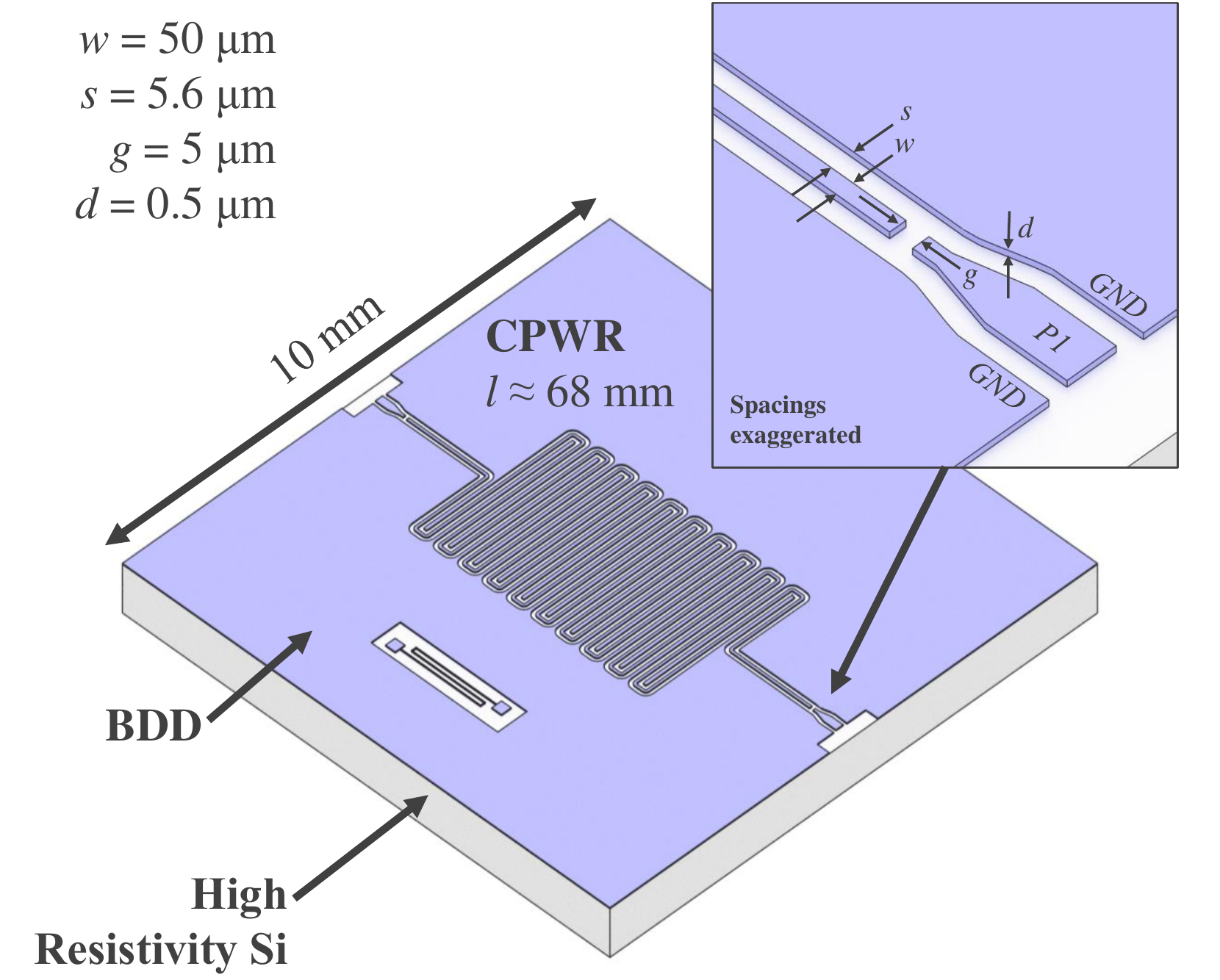}
	\caption{3D render schematic of the device layout (inset not to scale). `P1' denotes port for the microwave signal lines and `GND' denotes the ground planes. \label{fig:dev}}
\end{figure}
A microwave CPR is a well understood structure, with a designed resonant frequency related to the electrical length of the transmission line and its phase velocity. For a half-wavelength ($\lambda/2$) resonator with an infinite number of integer modes, the analytical frequencies are given by\cite{Goppl2008,Hahnle2020a}:
\begin{equation}
	f_r = \frac{nc}{2l \sqrt{\muEff\epsEff}} = \frac{n}{2l\sqrt{L_{\rm{0}}C_{\rm{0}}}} \label{eq:fr}
\end{equation}
where $f_r$ is the resonant frequency in Hz, $c$ is the speed of light in m/s, $l$ is the length of the transmission line in m, $\lambda$ is the wavelength in m, $n$ is the integer mode number, $\muEff$ and $\epsEff$ are the effective relative permeability and permittivity of the medium, respectively and $L_{\rm{0}}$ and $C_{\rm{0}}$ denote the total inductance and capacitance per unit length of the structure in H/m and F/m, respectively. $\muEff$ is assumed unity and $\epsEff$ is the average relative permittivity of the air space region and the substrate ($\epsEff\approx[\varepsilon_{\rm{r,sub}}+1]/2$, where $\varepsilon_{\rm{r,sub}}$ is the relative permittivity of the substrate). In the absence of any superconductivity, the only contribution to the stored electromagnetic energy in the resonator is the per unit length inductance ($L_g$) and capacitance ($C_g$) of the structure, or the `geometric' contributions, which through conformal mapping are estimated as\cite{Watanabe1994}:

\begin{align}
	L_g & = \frac{\mu_0}{4} \frac{K\left(k'\right)}{K\left(k\right)} \label{eq:Lg}\\
	C_g & = 4\varepsilon_0\epsEff \frac{K\left(k\right)}{K\left(k'\right)} \label{eq:Cg} \\
	k &= \frac{w}{w+2s} \\
	k' &= \sqrt{1-k^2}
\end{align}
where $\mu_0$ and $\varepsilon_0$ are the permeability and permittivity of free space in H/m and F/m respectively, $K$ is the complete elliptical integral of the first kind, $w$ is the width of the centre conductor in m and $s$ is the gap between the centre conductor and the ground plane in m. For a superconductor however, as the temperature is varied and transitions from the normal state into the superconducting state, the inertia of the migrating Cooper pairs becomes non-negligible and $L_k$ adds to the total inductance ($L_0 = L_g + L_k(T)$)\cite{Watanabe1994}. This kinetic inductance is a function of the magnetic penetration depth:
\begin{equation} 
	L_k(T) = \mu_0\frac{\penD(T)^2}{dw}g(w,s,d) \label{eq:Lk}
\end{equation}
where $\penD(T)$ is the temperature dependent magnetic penetration depth, $d$ is the film thickness, $g$ is the structural dependent geometric factor\cite{Watanabe1994}. The shift in frequency caused by $L_k$, can be modelled as a change in the phase velocity by introducing an effective medium. In a simple model, $\muEff$ in (\ref{eq:fr}) can be modelled using (\ref{eq:Lg}), (\ref{eq:Cg}) and (\ref{eq:Lk}):
\begin{multline}
	\muEff(T) \approx \left[\frac{1}{4}\frac{K\left(k'\right)}{K\left(k\right)} + \frac{\lambda(T)^2}{dw}g(w,s,d)\right]\times4\frac{K\left(k\right)}{K\left(k'\right)} \label{eq:muEff}
\end{multline}
Note that in the absence of any kinetic inductance (i.e. $\lambda(T)=0$), $\muEff=1$. Thus, the difference between the measured frequency ($f_m(T)$) and the analytical frequency ($f_r$) in (\ref{eq:fr}) can be attributed to $L_k$ and $\lambda(T)$. Additionally, the quality factor of the CPR can be related to the effective surface resistance using the following\cite{Yoshida1995}:
\begin{equation} 
	R_{s,\rm{eff}}(T) = \mu_0\lambda(T) \frac{\pi f_m(T)}{Q_U(T)} \frac{L_k+L_g}{L_k}
	\label{eq:Rs}
\end{equation}
where $R_{s,\rm{eff}}(T)$ is the temperature dependent surface resistance in ohms and $Q_U$ is the unloaded quality factor.

The CPR can be terminated with either short or open circuits, to which microwaves are coupled to the resonator using inductive or capacitive coupling, respectively. There are numerous methods of probing the resonance including using a feed-line or transmission directly through the resonator. The feed line approach `taps' off the microwaves to the CPR and is extremely advantageous for multiplexing multiple resonators on the same chip\cite{Porch2005}. However, for a sensitive readout, the coupling reactance needs to be sufficiently large in order to observe the resonance minimum. The coupling structures further load the resonator which results in a measured frequency and $Q$ factor containing contributions from both the antennas and the intrinsic resonance\cite{Goppl2008}. In the direct transmission approach, power transmission only occurs at resonance when the device superconducts and typically only one device can be measured. However assuming a high dynamic range can be measured through the instrumentation, extremely weak coupling can be achieved such that the measured properties are approximated to the intrinsic resonance\cite{JavaheriRahim2016}. For simple material investigations, the latter approach is favourable, as to minimise unknown variability caused by the coupling structures for different materials. The CPR device used in this study is an open circuit coupled $\lambda/2$ resonator shown in Figure \ref{fig:dev}, whereby a meandering coplanar waveguide allows an electrical length of approximately $l\approx$ 68.6 mm to be achieved on a $10\times10$ mm$^2$ substrate. Using (\ref{eq:fr}) and assuming $L_k=0$, $f_r \approx 867$ MHz). 

\section{Model}\label{sec:model}
A simple 3D FEM solution was developed in COMSOL Multiphysics\textregistered{} to check the resonant modes of the CPR meander design. Briefly, a 2D work-plane is used to define the 10$\times$10 mm$^2$ CPR lithography pattern. The work plane is then sandwiched between two 10$\times$10$\times$0.5 mm$^3$ domains, with the top defined as air ($\varepsilon_{\rm{r}}=1$) and the bottom as Si ($\varepsilon_{\rm{r}}=11.7$, $\tan\delta=2\times10^{-4}$\cite{Krupka2007}). The BDD layers are simulated as infinitely thin perfect electric conductors (PEC) with two multi-element lumped ports defined between the centre conductor pads and the ground planes. External boundaries are defined as PEC boundaries. The model is first run in the absence of any kinetic inductance ($\muEff=1$) to obtain the unperturbed resonance. Subsequently, the kinetic inductance is introduced ($L_k = 0.1 $ to \microH{1.5}/m) by modelling all domains with the effective magnetic permeability defined in (\ref{eq:muEff}). An electromagnetic eigenfrequency study was used to estimate the resonant mode frequencies and a frequency domain study was used to estimate the S-parameters. 

\begin{figure}[t!]
	\centering
	\includegraphics[width=0.45\textwidth]{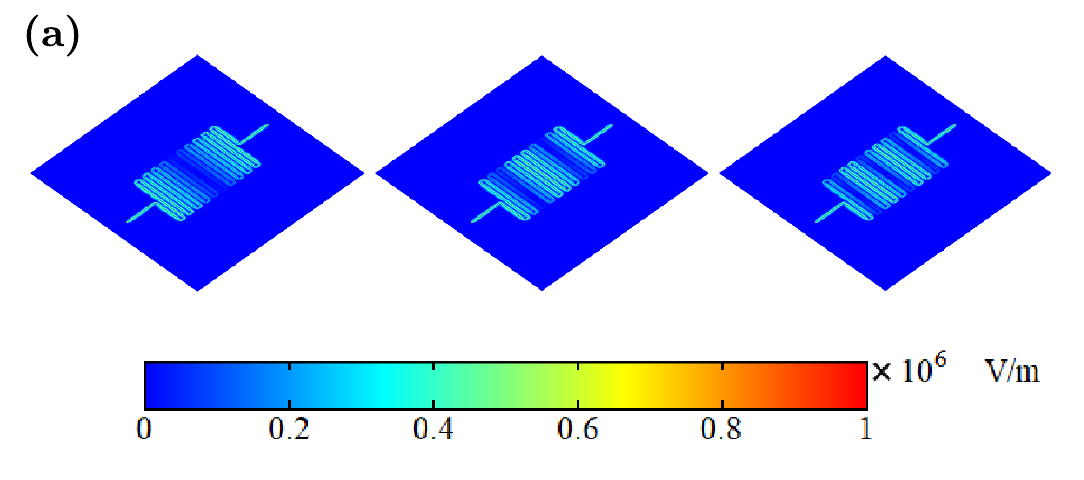} \\
	\includegraphics[width=0.45\textwidth]{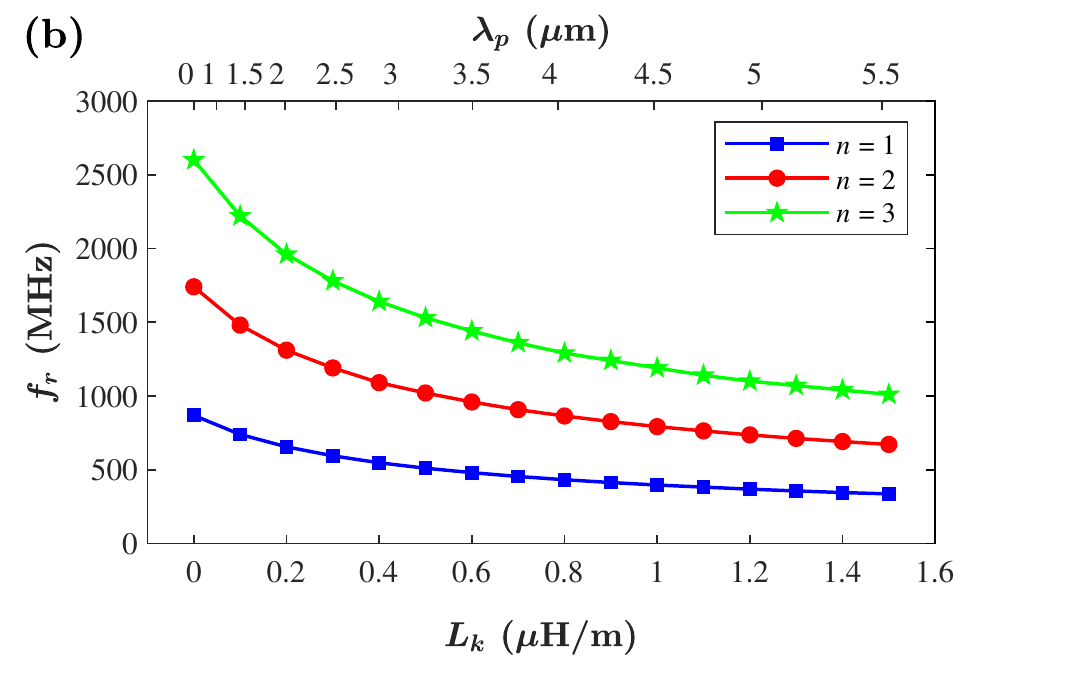}
	\caption{FEM results of (a) the electric (E) field distribution for $n=1,2$ and 3 and (b) the calculated resonant frequency as a function of $L_k$ (bottom axes) and $\penD$ (top axes) using the effective medium model. \label{fig:fem}}
\end{figure}

The FEM results are given in Figure \ref{fig:fem}, showing the calculated eigenfrequencies in the gigahertz range. The field distributions clearly show the first, second and third harmonics (868, 1736 and 2604 MHz). After introducing $L_k$ using the effective medium in (\ref{eq:muEff}), the fundamental resonant frequency decreases to as low as 336 MHz for the fundamental mode when $L_k$ is as large as \microH{1.5}/m or $\penD\approx$ \micron{3.1} for this structure with similar fractional frequency shifts are observed for other modes. 

\section{Method}\label{sec:method}
B-NCD films were grown in a Seki Diamond systems AX 6500 series MPCVD system. High resistivity  float-zone Si substrates ($\diameter$ = 2", $t$ = \micron{500}) were first seeded using a nanodiamond colloid suspension\cite{Mandal2021a,Williams2008} and grown at 3.5 kW at 40 Torr with a pyrometer measured substrate temperature of 755 $^\circ$C for 3 hours. Deposition was achieved in a gas mixture of CH$_4$, trimethylboron (TMB) dilute in H$_2$ in a total flow rate of 500 sccm (3\% CH4 and B/C ratio $\sim$12,800 ppm)\cite{Mandal2019b}. 

The CPR device was patterned using a standard photolithography process (solvent cleaning and AZ nLof 2020 resist recipe with an adhesion promoter). A 100 nm nickel mask was evaporated onto the B-NCD using an Edwards 306 physical vapour deposition system with subsequent lift off. The samples were then etched in an O$_2$/SF$_6$ inductively coupled plasma (ICP) using an Oxford Instruments PlasmaPro 100 Cobra system. Scanning electron microscopy (SEM) images were obtained using a Hitachi SU8200 (\SI{10}{\kilo\volt} at \SI{10}{\micro\ampere}) and Raman spectra was obtained using a Horiba LabRAM HR Evolution ($\lambda=$ \SI{532}{\nano\metre}). Temperature dependent resistance measurements were also carried out on a separate piece of the same wafer using a Van der Pauw configuration in a Quantum Design physical property measurement system from 2 to 300 K .

The CPR device was cooled using a custom-made cooling platform utilising a Cryomech PT-407-RM pre cooler and two Chase Research Cryogenics sorption fridges. The device measurement is achieved using a Rhode \& Schwarz ZNB vector network analyser (VNA) connected to an RF chain of attenuators with the return signal boosted using low noise amplifiers from (Arizona State University and L3 Narda-MITEQ). Power transmission data was obtained using the VNA (magnitude $|S_{21}|^2$ and phase) and resonator parameter fitting was carried out using a MATLAB script.


\begin{figure}[t!]
	\centering\noindent
	\includegraphics[trim={9.5cm 5cm 9.5cm 5cm},clip, width=0.45\textwidth]{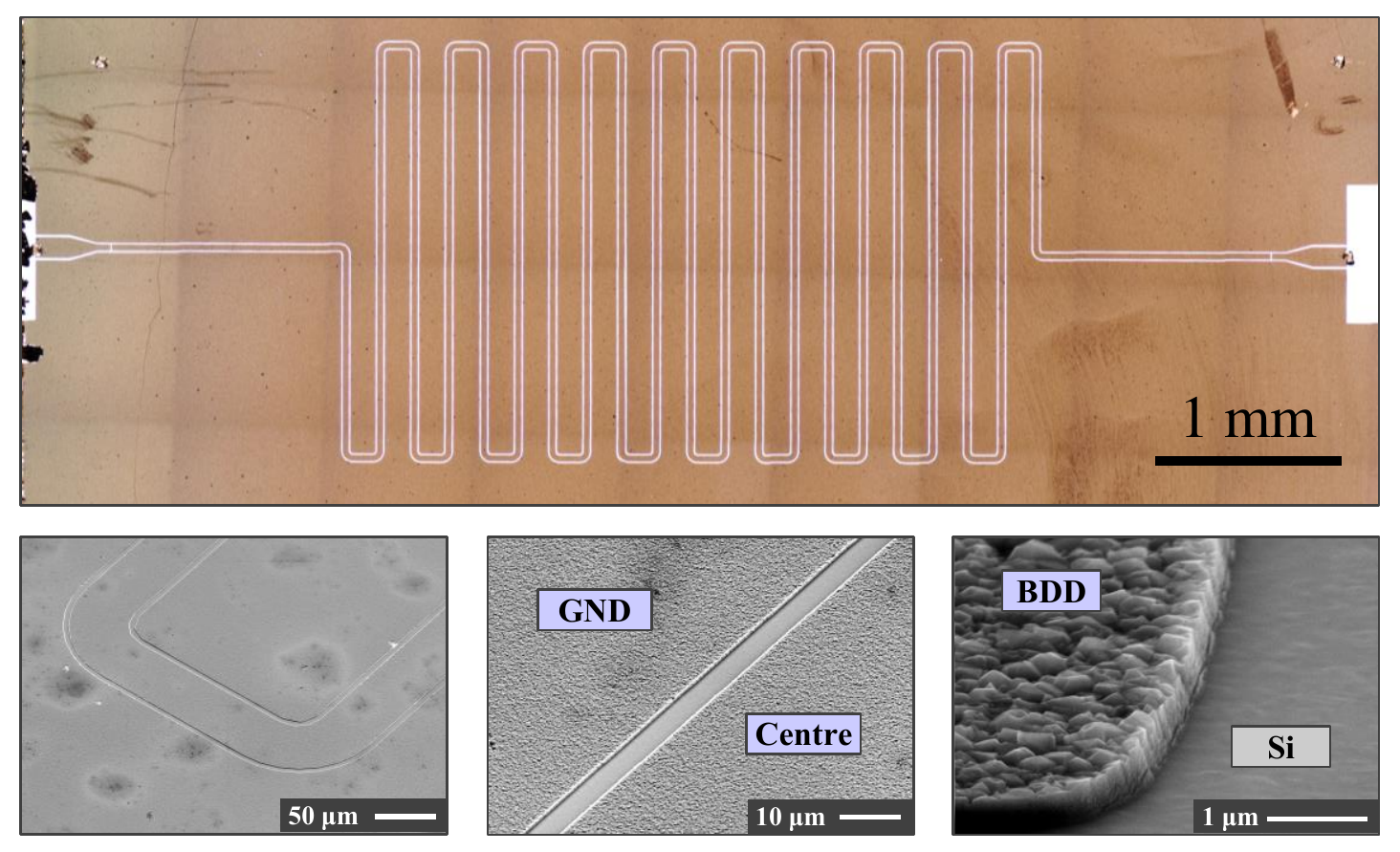}
	\caption{Microscope and SEM images of the B-NCD CPR device. \label{fig:sem}}
\end{figure}
\begin{figure}[t!]
	\centering
	\includegraphics[width=0.45\textwidth]{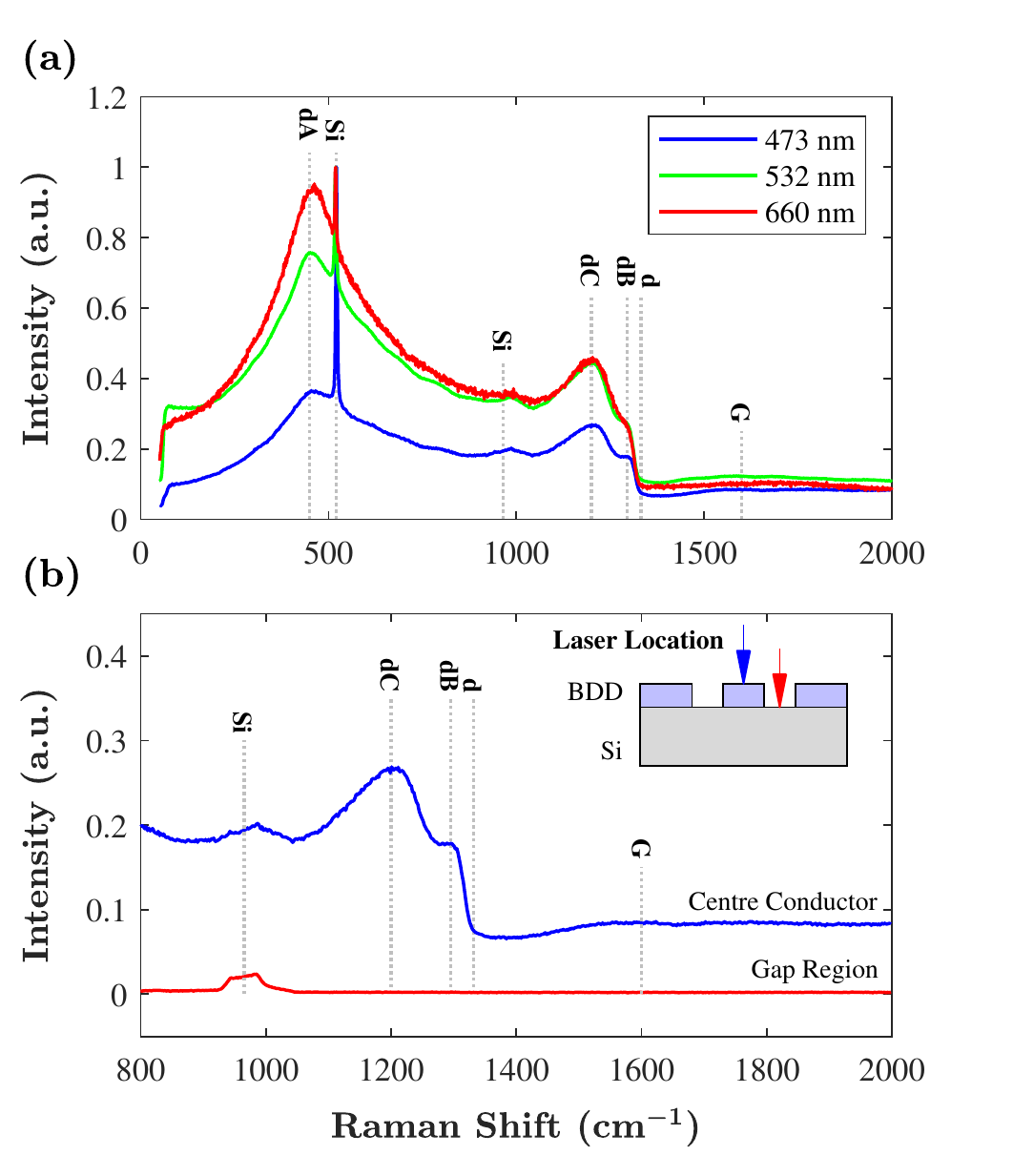}
	\caption{Raman spectra of B-NCD CPR device: (a) Wide survey spectra of the centre conductor using 473, 532 and 660 nm and excitation wavelengths and (b) shows a magnified region situated around the diamond `d' band.\label{fig:raman}}
\end{figure}
\section{Results}\label{sec:exp}
Microscope and SEM images of the CPR device after ICP etching are shown in Figure \ref{fig:sem}, showing the uninterrupted length of the transmission line structure with the  capacitive coupling gaps at the edges of the sample. The SEM images show that in the gap regions, the B-NCD has been successfully etched which was also corroborated using a probe station measurement. Raman spectra is given in Figure \ref{fig:raman}, showing the first and second order bands of Si at $\sim$520 \cm{} and $\sim$965 \cm{}, respectively\cite{Prawer2004a, Sedov2019,Cuenca2022} and numerous signatures typically associated with BDD at high dopant concentrations. The B-NCD bands are labelled at $\sim$450 \cm{} (dA), $\sim$1200 \cm{} (dC), $\sim$1295 \cm{} (dB), 1332 \cm{} (d) and the wide region at $\sim$1500-1680 \cm{} (G)\cite{Mortet2017b}. Figure \ref{fig:raman}(a) shows that there is minimal variation in the spectra as a function of the laser excitation wavelength, where typically the G band becomes prominent at high laser wavelengths\cite{Prawer2004a}. The magnified region given in Figure \ref{fig:raman}(b) shows that the characteristic zone centre phonon line at 1332 \cm{} (line `d') is redshifted to lower Raman shifts which is a typical signature of highly boron doped nan-ocrystalline  films\cite{Mortet2017b,Mandal2019b,Zibrov2018,Nagasaka2016,Dubrovinskaia2006}. The low intensity G band for all excitation wavelengths implies that while the film is nanocrystalline with considerable grain boundaries, the non-diamond carbon concentration is low. In the gap regions, there are no B-NCD signatures and only those from the underlying Si.

\begin{figure}[t!]
	\centering
	\includegraphics[width=0.45\textwidth]{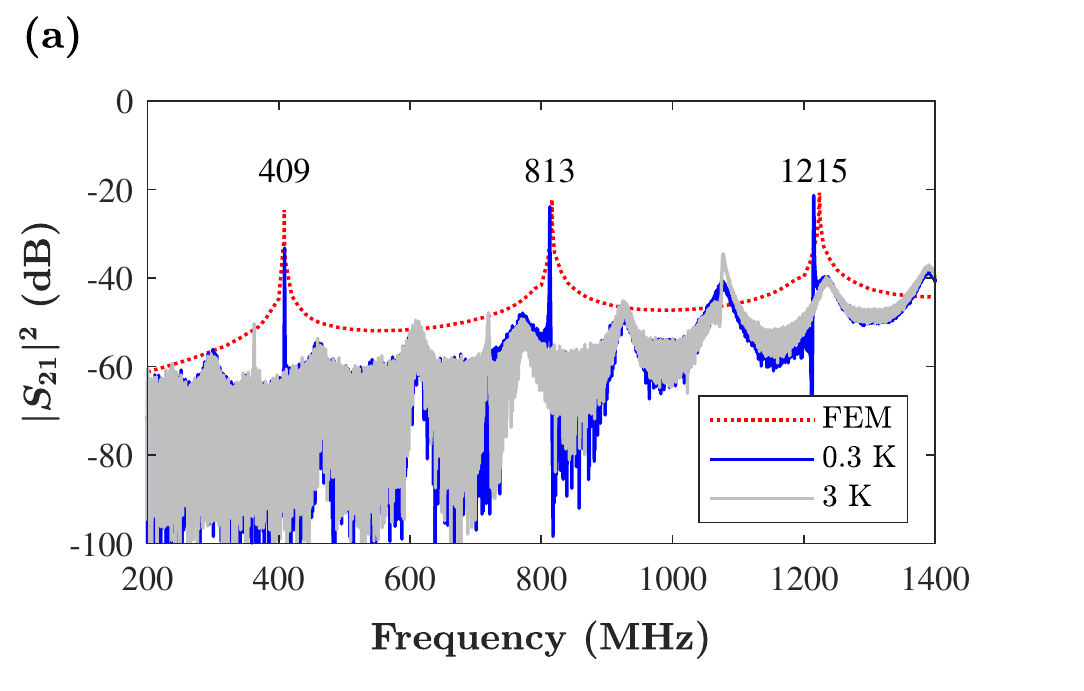}
	\includegraphics[width=0.45\textwidth]{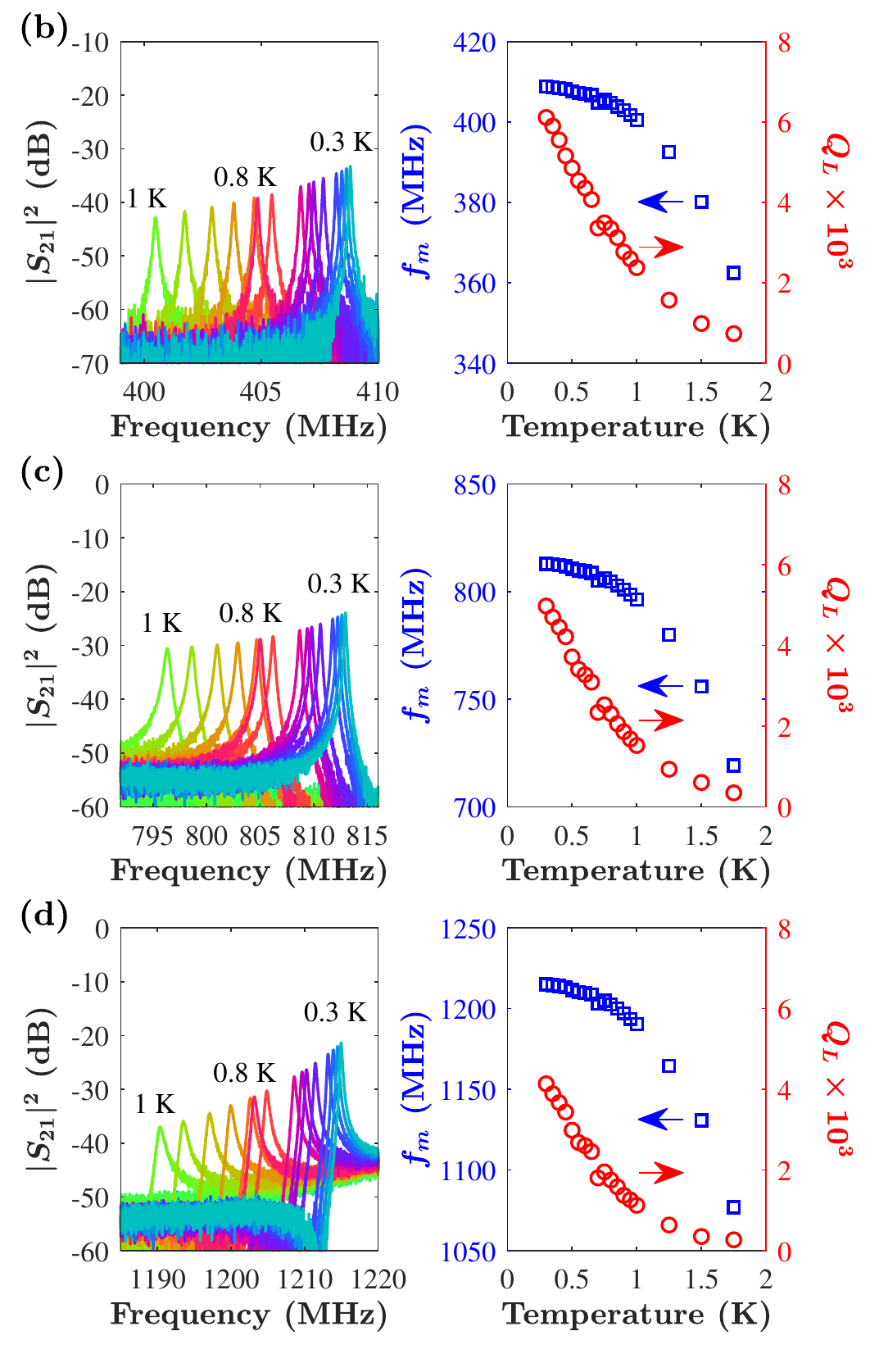}
	\caption{Power transmission spectra of the B-NCD CPR at ($P_{\textrm{in}} =$ 0 dBm with 20 dB attenuation). (a) shows the wide $|S_{21}|^2$ spectra at 3 K (grey) and 0.3 K (blue) with an FEM result for $L_k\approx\microH{0.93}$/m (red dotted). (b) to (c) show magnified views of the 3 modes from 1 K to 0.3 K and the fitted $f_m$ and $Q_L$.\label{fig:mw}}
\end{figure}

The power transmission characteristics of the CPR are shown in Figure \ref{fig:mw}. The broadband spectra in Figure \ref{fig:mw} (a) shows that at 3 K, the device fails to resonate and only a background noise floor at $\approx-60$ dB is measurable. As $T$ is lowered towards 300 mK, sharp resonances emerge at approximately 409, 813 and 1215 MHz, much lower than the designed frequencies of 868, 1736 and 2604 MHz, respectively. Using an effective kinetic inductance of $L_k \approx$ \microH{0.93}/m, the FEM frequency domain $|S_{21}|^2$ traces demonstrate a plausible correlation with the experimental measurements. Additionally, the peak power transmission is low, demonstrating weak coupling. Magnified views of each resonance as a function of $T$ are shown in Figure \ref{fig:mw} (b) to (d). Each of the spectra were fitted to a Lorentzian function to extract the measured resonant frequency and the loaded quality factor\cite{Cuenca2015b}:
\begin{equation}
	|S_{21}|^2 = \frac{P_0}{1+4Q_L^2\left(\frac{f-f_m}{f_m}\right)^2}
\end{equation}
where $P_0$ is the peak power at resonance, $Q_L=Q_U(1-P_0)$ is the loaded quality factor and $f$ and $f_m$ are the frequency and the measured resonant frequency, respectively in Hz. Note that for all resonances, the peak transmitted power $P_0\ll0$ dB such that $Q_L\approx Q_U$ where the resonator is weakly coupled, thus mitigating the need for any de-embedding of the coupling capacitance. This was further confirmed using power sweeps whereby minimal shift in the resonant frequency was observed (not shown). The resonant frequency increases as $T$ decreases and is observed to saturate as  $T<300$ mK with minimal back-bending. Nominal quality factors range from 4,000 to 6,000 at 300 mK, however, saturation was not observed owing to the limited $T$ range. It is quite possible that the intrinsic Q was actually much higher towards $T\rightarrow0$ K. 

\begin{figure}[t!]
	\centering
	\includegraphics[width=0.45\textwidth]{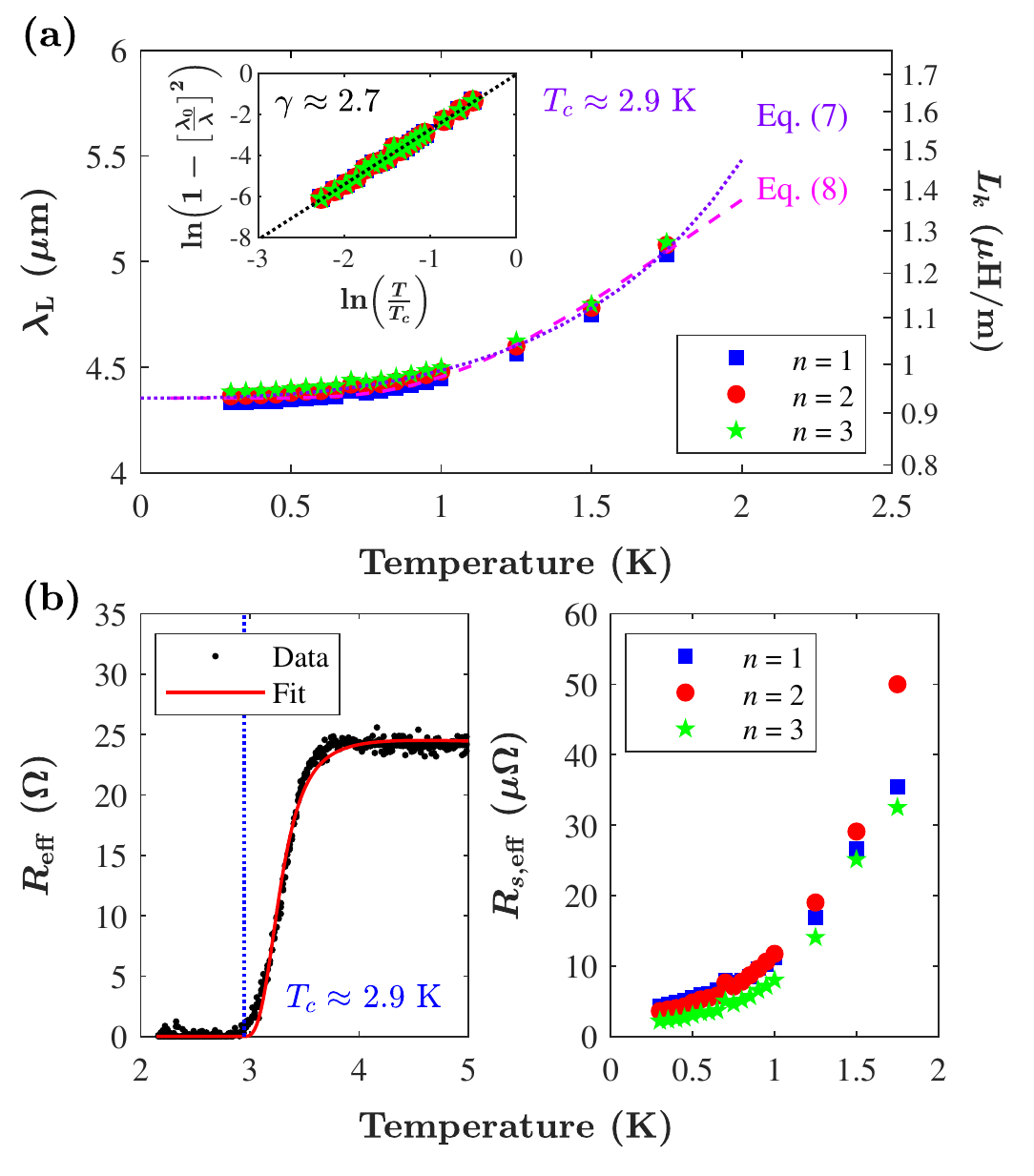}
	\caption{Calculated superconducting properties of B-NCD CPR device. (a) shows the per unit length $\penD$ and $L_k$ and (b) shows the van der Pauw $R(T)$ (left) and the calculated $R_{s,\textrm{eff}}(T)$ (right). Identification of $T_c$ is denoted by the divergence in resistance (blue dotted) \label{fig:lk}}
\end{figure}

Using (\ref{eq:fr}) and (\ref{eq:muEff}), the estimated  $L_k$ and $\penD$ are shown in Figure \ref{fig:lk}(a). Based on these frequency shifts, the extracted kinetic inductance is $\approx$ \microH{1.25}/m with an associated penetration depth of $\approx$ \micron{5} at 1.75 K and decreases to $\approx$ \microH{0.93}/m with an associated penetration depth $\approx$ \micron{4.3} at 300 mK. Since all harmonics are linked by $n$, similar values were obtained across the modes. To estimate the zero temperature penetration depth, temperature dependent models are required, with one of the most typical approaches being the empirical two fluid model as presented by Gorter and Casimir (GC):
\begin{equation}
	\penD(T) = \frac{\lambda_0}{\left[1-\left(\frac{T}{T_c}\right)^{\gamma}\right]^{\frac{1}{2}}}
	\label{eq:gc}
\end{equation}
where $\lambda_0$ is the penetration depth at 0 K and $\gamma=4$, though experimental measurements rarely exhibit this exponent depending on the type of superconductor\cite{Prozorov2006}. $T_c$ is estimated using the van der Pauw measurement in Figure \ref{fig:lk} (b), giving $T_c\approx 2.9$ K which is in agreement with B-NCD films grown in the same reactor at similar conditions ($T_c \approx$ 2.4 to 4.5 K)\cite{Klemencic2017,Klemencic2017b,Manifold2021}. Using this value and (\ref{eq:gc}), the 0 K extrapolated values are given in Table \ref{tab:lk} for $\gamma=2.7$. Similarly, the temperature dependent penetration depth can also be related to the superconducting gap energy through the BCS low-temperature model:
\begin{equation}
	\lambda_{L}(T) = \lambda_0\sqrt{\frac{\pi\Delta_0}{2k_BT}}\exp{\left(\frac{-\Delta_0}{k_BT}\right)}+\lambda_0
\label{eq:gap}
\end{equation}
where $\Delta_0$ is the superconducting gap energy at 0 K and $k_B$ is the Boltzmann constant. Figure \ref{fig:lk}(a) shows the fit to $\penD(T)$ using the same $\lambda_0$ as (\ref{eq:gc}), yielding $\Delta_0 \approx$ \microEV{895}. Finally, the surface resistance values obtained using (\ref{eq:Rs}) are given in Figure \ref{fig:lk}(b), demonstrating that  $R_{s,\rm{eff}}(T)$ decreases from 30 to \microO{50} at 1.75 K to 2 to \microO{4} at 300 mK.

\begin{table}[t!]\small\centering
\caption{\label{tab:lk}Calculated zero temperature kinetic inductance per unit length and per square and penetration depth based on the CPR effective medium model\footnotesize{$^{\textrm{a}}$} }
\begin{tabular}{cccc}
\hline
Mode, $n$ & $L_{k,0}$ (\microH{}/m) &$L_{k\square}$ (pH/$\square$)\footnotesize{$^{\textrm{b}}$} & $\lambda_0$ (\micron{})\\
{}	&	$\pm0.006$ & $\pm4$ & $\pm0.02$ \\
\hline
1 & 0.92 & 671& 4.33\\
2 & 0.93 & 681 & 4.36\\
3 & 0.94& 687 & 4.38\\
\hline
\end{tabular} \\ \raggedright
\footnotesize{$^{\textrm{a}}$ Error assumes a conservative measurement uncertainty in $f_m$ of $\pm 1$ MHz} \\
\footnotesize{$^{\textrm{b}}$ Calculated by assuming $L_{k\square} = L_{k,0} \times w/l$} \\
\end{table}

\begin{table}[t!]\small\centering
\caption{\label{tab:Rs}Calculated zero temperature surface resistance and sheet resistance based on the  CPR effective medium model\footnotesize{$^{\textrm{a}}$} }
\begin{tabular}{cccc}
\hline
Mode, $n$ & $R_{s,\rm{eff}}$ ($\mu\Omega$/m) &$R_{s,\square}$ (\SI{}{\nano\ohm}$/\square$)\footnotesize{$^{\textrm{b}}$}  \\
{}	&	$\pm0.01$ & $\pm0.007$ \\
\hline
1 & 4.4 & 3.2\\
2 & 3.6 & 2.6\\
3 & 2.2 & 1.6\\
\hline
\end{tabular} \\ \raggedright
\footnotesize{$^{\textrm{a}}$ Error assumes a conservative measurement uncertainty in $Q_U$ of $\pm10$.} \\
\footnotesize{$^{\textrm{b}}$ Calculated by $R_{s,\square}=R_{s,\rm{eff}} \times w/l$ }
\end{table}

\section{Discussion}\label{sec:disc}
The results presented here suggest that using the B-NCD films have a considerably large kinetic inductance and large penetration depths, far greater than the theoretical London penetration depths of BDD $\lambda_0 = (\sfrac{m_e}{\mu_0ne^2})^{1/2}\approx 168$ nm where $m_e$ and $e$ are the electron mass and charge, respectively and $n>10^{21}$cm$^{-3}$. The penetration depths are more consistent with those experimentally found in thin B-NCD films on Si substrates at a similar temperature ranges; $\penD \approx$ 2 to \micron{4}\cite{Oripov2021}, however, less consistent with much thicker films or those measured on single crystal diamond substrates or B-SCD; $\penD \approx$ 0.2 to \micron{1}\cite{Ortolani2006,Winzer2005}. The results support the evidence that the granularity of the films plays a large role in increasing the penetration depth, and therefore its sensitivity for microwave kinetic inductance based devices. The laminar model presented by Hylton and Beasley considers the effect of large and small grain sizes on the effective penetration depth, and demonstrate that polycrystalline materials in general have much larger effective penetration depths than single crystal owing to the grain boundary separation of superconducting grains\cite{Hylton1989}. 

The temperature dependence of $\penD$ as described by (\ref{eq:gap}) yields a $\gamma$ exponent that tends towards that of an s-wave superconductor, however, the exponent likely due to the heavily disordered and granular nature of B-NCD\cite{Prozorov2006}. The extracted superconducting conducting gap is large, however, is of similar magnitude to those reported by other measurement techniques.  Ishizaka et al. obtained a gap of $\Delta_0\approx$\microEV{780} by laser excited photoemission spectroscopy, Sac\'{e}p\'{e} et al. obtained $\Delta_0\approx\microEV{285}$ by tunnelling spectroscopy and finally, Oripov et al. obtained a gap $\Delta_0\approx\microEV{924}$ through microwave parallel plate resonator (PPR) \cite{Ishizaka2007,Sacepe2006}. 

The calculated surface resistance is very low at 2-\microO{4}, and comparable to measurements of YBCO at similar frequencies although at much lower temperatures\cite{Wang2007}. Additionally, the calculated values here are much lower than those reported previously for B-NCD films in this frequency range, as is apparent by the higher observed quality factors\cite{Oripov2021}. Since the presented method and that of previous approaches using microwave PPR are both capable of measuring to similarly low values of $R_s$\cite{Yoshida1995,Taber1990}, the discrepancy may lie in the material deposition approach, i.e. by either microwave plasma assisted chemical vapour deposition (MP-CVD) or hot filament chemical vapour deposition (HF-CVD). Grain boundaries are typically apparent in B-NCD films, irrespective of the deposition methodology, however, for films grown using MP-CVD, very low concentrations of non-diamond carbon are typically observed by the low G band in Raman spectroscopy, in contrast to those by HF-CVD\cite{Mortet2017b,Manifold2021,Kumar2017a}. This result suggests that non-diamond carbon microwave losses are potentially significant at low temperatures. Regardless, the measured quality factor of this device for B-NCD is still fairly low when compared to other material technologies based on aluminium\cite{Zhang2019b,Kalacheva2020} and niobium\cite{Hornibrook2012} at $Q_U>10^4$. The cause of this is likely the substrate to which the B-NCD is deposited on; using the FEM model to reduce the loss tangent to zero yields Q factors $>10^5$. High resistivity float-zone Si substrates are known to have additional losses at low $T$ owing to potential hopping conduction\cite{Krupka2007}. Additionally, boron diffusion into the Si is likely to occur during the plasma assisted growth process resulting in an additional dielectric loss from the substrate\cite{Shao2003,Mirabella2013,Shenai1992}. Further improvements could potentially be made using substrates to stop boron diffusion, or more robust dielectrics that are less affected by microwave H$_2$ plasmas which are out of scope of this work.

\section{Conclusion}
This work demonstrates a measurement of the seemingly large penetration depth and low surface resistance of B-NCD using a microwave CPR approach with an effective medium model. Using a simple meander design with a weakly coupled power transmission technique, the intrinsic superconducting properties of thin B-NCD films can be obtained. This work demonstrates that coplanar microwave devices made from B-NCD can be potentially used for highly sensitive kinetic inductance detectors.

\section{Acknowledgements}
This project has been supported by Engineering and Physical Sciences Research Council (EPSRC) under the GaN-DaME program grant (EP/P00945X/1) and the ``A Diamond Bridge to Phase Slip Physics'' grant (EP/V048457/1 ). This project has been supported by the European Research Council (ERC) Consolidator Grant under the SUPERNEMS Project (647471). SEM was carried out in the clean-room of the ERDF-funded Institute for Compound Semiconductors (ICS) at Cardiff University.

\bibliographystyle{elsarticle-num-nourl.bst}
\bibliography{\jobname}

\begin{thebibliography}{10}
\expandafter\ifx\csname url\endcsname\relax
  \def\url#1{\texttt{#1}}\fi
\expandafter\ifx\csname urlprefix\endcsname\relax\def\urlprefix{URL }\fi
\expandafter\ifx\csname href\endcsname\relax
  \def\href#1#2{#2} \def\path#1{#1}\fi

\bibitem{Gajewski2009}
W.~Gajewski, P.~Achatz, O.~A. Williams, K.~Haenen, E.~Bustarret, M.~Stutzmann,
  et~al., {Electronic and optical properties of boron-doped nanocrystalline
  diamond films}, Physical Review B - Condensed Matter and Materials Physics
  79~(4) (2009) 1--14.
\newblock \href {https://doi.org/10.1103/PhysRevB.79.045206}
  {\path{doi:10.1103/PhysRevB.79.045206}}.

\bibitem{Cobb2018}
S.~J. Cobb, Z.~J. Ayres, J.~V. Macpherson,
  \href{http://10.0.4.122/annurev-anchem-061417-010107
  https://dx.doi.org/10.1146/annurev-anchem-061417-010107}{{Boron Doped
  Diamond: A Designer Electrode Material for the Twenty-First Century}}, Annual
  Review of Analytical Chemistry 11~(1) (2018) 463--484.
\newblock \href {https://doi.org/10.1146/annurev-anchem-061417-010107}
  {\path{doi:10.1146/annurev-anchem-061417-010107}}.

\bibitem{Ekimov2004}
E.~A. Ekimov, V.~A. Sidorov, E.~D. Bauer, N.~N. Mel'nik, N.~J. Curro, J.~D.
  Thompson, et~al.,
  \href{http://www.nature.com/articles/nature02449}{{Superconductivity in
  diamond}}, Nature 428~(6982) (2004) 542--545.
\newblock \href {https://doi.org/10.1038/nature02449}
  {\path{doi:10.1038/nature02449}}.

\bibitem{Winzer2005}
K.~Winzer, D.~Bogdanov, C.~Wild, {Electronic properties of boron-doped diamond
  on the border between the normal and the superconducting state}, Physica C:
  Superconductivity and its Applications 432~(1-2) (2005) 65--70.
\newblock \href {https://doi.org/10.1016/j.physc.2005.07.011}
  {\path{doi:10.1016/j.physc.2005.07.011}}.

\bibitem{Takano2004}
Y.~Takano, M.~Nagao, I.~Sakaguchi, M.~Tachiki, T.~Hatano, K.~Kobayashi, et~al.,
  {Superconductivity in diamond thin films well above liquid helium
  temperature}, Applied Physics Letters 85~(14) (2004) 2851--2853.
\newblock \href {https://doi.org/10.1063/1.1802389}
  {\path{doi:10.1063/1.1802389}}.

\bibitem{Bustarret2004}
E.~Bustarret, J.~Kacmar{\v{c}}ik, C.~Marcenat, E.~Gheeraert, C.~Cytermann,
  J.~Marcus, et~al., {Dependence of the superconducting transition temperature
  on the doping level in single-crystalline diamond films}, Physical Review
  Letters 93~(23) (2004) 2--5.
\newblock \href {https://doi.org/10.1103/PhysRevLett.93.237005}
  {\path{doi:10.1103/PhysRevLett.93.237005}}.

\bibitem{Mandal2011}
S.~Mandal, T.~Bautze, O.~A. Williams, C.~Naud, {\'{E}}.~Bustarret,
  F.~Omn{\`{e}}s, et~al., \href{http://aip.scitation.org/doi/10.1063/1.3561743
  https://pubs.acs.org/doi/10.1021/nn2018396}{{The Diamond Superconducting
  Quantum Interference Device}}, ACS Nano 5~(9) (2011) 7144--7148.
\newblock \href {https://doi.org/10.1021/nn2018396}
  {\path{doi:10.1021/nn2018396}}.

\bibitem{Kageura2019}
T.~Kageura, M.~Hideko, I.~Tsuyuzaki, A.~Morishita, A.~Kawano, Y.~Sasama,
  et~al., {Single-crystalline boron-doped diamond superconducting quantum
  interference devices with regrowth-induced step edge structure}, Scientific
  Reports 9~(1) (2019) 2--8.
\newblock \href {https://doi.org/10.1038/s41598-019-51596-w}
  {\path{doi:10.1038/s41598-019-51596-w}}.

\bibitem{Klemencic2017}
G.~M. Klemencic, S.~Mandal, J.~M. Werrell, S.~R. Giblin, O.~A. Williams,
  {Superconductivity in planarised nanocrystalline diamond films}, Science and
  Technology of Advanced Materials 18~(1) (2017) 239--244.
\newblock \href {https://doi.org/10.1080/14686996.2017.1286223}
  {\path{doi:10.1080/14686996.2017.1286223}}.

\bibitem{Mandal2010b}
S.~Mandal, C.~Naud, O.~A. Williams, E.~Bustarret, F.~Omn{\`{e}}s,
  P.~Rodi{\`{e}}re, et~al., {Nanostructures made from superconducting
  boron-doped diamond}, Nanotechnology 21~(19) (2010).
\newblock \href {https://doi.org/10.1088/0957-4484/21/19/195303}
  {\path{doi:10.1088/0957-4484/21/19/195303}}.

\bibitem{Mandal2010a}
S.~Mandal, C.~Naud, O.~A. Williams, {\'{E}}.~Bustarret, F.~Omn{\`{e}}s,
  P.~Rodi{\`{e}}re, et~al.,
  \href{https://onlinelibrary.wiley.com/doi/10.1002/pssa.201000008}{{Detailed
  study of superconductivity in nanostructured nanocrystalline boron doped
  diamond thin films}}, physica status solidi (a) 207~(9) (2010) 2017--2022.
\newblock \href {https://doi.org/10.1002/pssa.201000008}
  {\path{doi:10.1002/pssa.201000008}}.

\bibitem{Ortolani2006}
M.~Ortolani, S.~Lupi, L.~Baldassarre, U.~Schade, P.~Calvani, Y.~Takano, et~al.,
  \href{https://link.aps.org/doi/10.1103/PhysRevLett.97.097002}{{Low-Energy
  Electrodynamics of Superconducting Diamond}}, Physical Review Letters 97~(9)
  (2006) 097002.
\newblock \href {https://doi.org/10.1103/PhysRevLett.97.097002}
  {\path{doi:10.1103/PhysRevLett.97.097002}}.

\bibitem{Oripov2021}
B.~Oripov, D.~Kumar, C.~Garcia, P.~Hemmer, T.~Venkatesan, M.~S. {Ramachandra
  Rao}, et~al., \href{https://aip.scitation.org/doi/10.1063/5.0051227}{{Large
  microwave inductance of granular boron-doped diamond superconducting films}},
  Applied Physics Letters 118~(24) (2021) 242601.
\newblock \href {http://arxiv.org/abs/2103.14738} {\path{arXiv:2103.14738}},
  \href {https://doi.org/10.1063/5.0051227} {\path{doi:10.1063/5.0051227}}.

\bibitem{McLean1962}
W.~L. McLean, {Superconducting penetration depth measurements in aluminium at
  175 Mc/s}, Proceedings of the Physical Society 79~(3) (1962) 572--585.
\newblock \href {https://doi.org/10.1088/0370-1328/79/3/314}
  {\path{doi:10.1088/0370-1328/79/3/314}}.

\bibitem{Cohen1968a}
R.~W. Cohen, B.~Abeles,
  \href{https://link.aps.org/doi/10.1103/PhysRev.168.444}{{Superconductivity in
  Granular Aluminum Films}}, Physical Review 168~(2) (1968) 444--450.
\newblock \href {https://doi.org/10.1103/PhysRev.168.444}
  {\path{doi:10.1103/PhysRev.168.444}}.

\bibitem{Watanabe1994}
K.~Watanabe, K.~Yoshida, T.~Aoki, {Kinetic inductance of superconducting
  coplanar waveguides}, Japanese Journal of Applied Physics 33~(10R) (1994)
  570--572.
\newblock \href {https://doi.org/10.1143/JJAP.33.5708}
  {\path{doi:10.1143/JJAP.33.5708}}.

\bibitem{Porch2005}
A.~Porch, P.~Mauskopf, S.~Doyle, C.~Dunscombe, {Calculation of the
  characteristics of coplanar resonators for kinetic inductance detectors},
  IEEE Transactions on Applied Superconductivity 15~(2 PART I) (2005) 552--555.
\newblock \href {https://doi.org/10.1109/TASC.2005.849916}
  {\path{doi:10.1109/TASC.2005.849916}}.

\bibitem{Doyle2008}
S.~Doyle, P.~Mauskopf, J.~Naylon, A.~Porch, C.~Dunscombe,
  \href{http://link.springer.com/10.1007/s10909-007-9685-2}{{Lumped Element
  Kinetic Inductance Detectors}}, Journal of Low Temperature Physics 151~(1-2)
  (2008) 530--536.
\newblock \href {https://doi.org/10.1007/s10909-007-9685-2}
  {\path{doi:10.1007/s10909-007-9685-2}}.

\bibitem{Noguchi2018}
T.~Noguchi, A.~Dominjon, Y.~Sekimoto,
  \href{http://ieeexplore.ieee.org/document/8302574/}{{Analysis of
  Characteristics of Al MKID Resonators}}, IEEE Transactions on Applied
  Superconductivity 28~(4) (2018) 1--6.
\newblock \href {http://arxiv.org/abs/1709.10421} {\path{arXiv:1709.10421}},
  \href {https://doi.org/10.1109/TASC.2018.2809615}
  {\path{doi:10.1109/TASC.2018.2809615}}.

\bibitem{Zhang2019a}
W.~Zhang, K.~Kalashnikov, W.-S. Lu, P.~Kamenov, T.~DiNapoli, M.~Gershenson,
  {Microresonators Fabricated from High-Kinetic-Inductance Aluminum Films},
  Physical Review Applied 11~(1) (2019) 011003.
\newblock \href {http://arxiv.org/abs/1807.00210} {\path{arXiv:1807.00210}},
  \href {https://doi.org/10.1103/PhysRevApplied.11.011003}
  {\path{doi:10.1103/PhysRevApplied.11.011003}}.

\bibitem{How1992b}
H.~How, R.~Seeds, C.~Vittoria, {Microwave Charactercistics of YBCO Coplanar
  Waveguide Resonator}, IEEE Transactions on Magnetics 28~(5) (1992)
  2217--2219.
\newblock \href {https://doi.org/10.1109/20.179448}
  {\path{doi:10.1109/20.179448}}.

\bibitem{Porch1995}
A.~Porch, M.~Lancaster, R.~Humphreys,
  \href{http://ieeexplore.ieee.org/document/348089/}{{The coplanar resonator
  technique for determining the surface impedance of
  YBa{\textless}sub{\textgreater}2{\textless}/sub{\textgreater}Cu{\textless}sub{\textgreater}3{\textless}/sub{\textgreater}O{\textless}sub{\textgreater}7-$\delta${\textless}/sub{\textgreater}
  thin films}}, IEEE Transactions on Microwave Theory and Techniques 43~(2)
  (1995) 306--314.
\newblock \href {https://doi.org/10.1109/22.348089}
  {\path{doi:10.1109/22.348089}}.

\bibitem{Yoshida1995}
K.~Yoshida, K.~Watanabe, T.~Kisu, K.~Enpuku, {Evaluation of Magnetic
  Penetration Depth and Surface Resistance of Superconducting Thin Films using
  Coplanar Waveguides}, IEEE Transactions on Applied Superconductivity 5~(2)
  (1995) 1979--1982.
\newblock \href {https://doi.org/10.1109/77.402973}
  {\path{doi:10.1109/77.402973}}.

\bibitem{Goppl2008}
M.~G{\"{o}}ppl, A.~Fragner, M.~Baur, R.~Bianchetti, S.~Filipp, J.~M. Fink,
  et~al., \href{http://aip.scitation.org/doi/10.1063/1.3010859}{{Coplanar
  waveguide resonators for circuit quantum electrodynamics}}, Journal of
  Applied Physics 104~(11) (2008) 113904.
\newblock \href {http://arxiv.org/abs/arXiv:0807.4094v1}
  {\path{arXiv:arXiv:0807.4094v1}}, \href {https://doi.org/10.1063/1.3010859}
  {\path{doi:10.1063/1.3010859}}.

\bibitem{Hahnle2020a}
S.~H{\"{a}}hnle, N.~V. Marrewijk, A.~Endo, K.~Karatsu, D.~J. Thoen,
  V.~Murugesan, et~al.,
  \href{http://aip.scitation.org/doi/10.1063/5.0005047}{{Suppression of
  radiation loss in high kinetic inductance superconducting co-planar
  waveguides}}, Applied Physics Letters 116~(18) (2020) 182601.
\newblock \href {http://arxiv.org/abs/2003.10241} {\path{arXiv:2003.10241}},
  \href {https://doi.org/10.1063/5.0005047} {\path{doi:10.1063/5.0005047}}.

\bibitem{JavaheriRahim2016}
M.~{Javaheri Rahim}, T.~Lehleiter, D.~Bothner, C.~Krellner, D.~Koelle,
  R.~Kleiner, et~al.,
  \href{https://iopscience.iop.org/article/10.1088/0022-3727/49/39/395501}{{Metallic
  coplanar resonators optimized for low-temperature measurements}}, Journal of
  Physics D: Applied Physics 49~(39) (2016) 395501.
\newblock \href {https://doi.org/10.1088/0022-3727/49/39/395501}
  {\path{doi:10.1088/0022-3727/49/39/395501}}.

\bibitem{Krupka2007}
J.~Krupka, J.~Breeze, N.~M. Alford, A.~E. Centeno, L.~Jensen, T.~Claussen,
  {Measurements of permittivity and dielectric loss tangent of high resistivity
  float zone silicon at microwave frequencies}, 16th International Conference
  on Microwaves, Radar and Wireless Communications, MIKON 2006 54~(11) (2007)
  3995--4001.
\newblock \href {https://doi.org/10.1109/MIKON.2006.4345377}
  {\path{doi:10.1109/MIKON.2006.4345377}}.

\bibitem{Mandal2021a}
S.~Mandal, {Nucleation of diamond films on heterogeneous substrates: a review},
  RSC Advances 11~(17) (2021) 10159--10182.
\newblock \href {https://doi.org/10.1039/d1ra00397f}
  {\path{doi:10.1039/d1ra00397f}}.

\bibitem{Williams2008}
O.~A. Williams, M.~Nesladek, M.~Daenen, S.~Michaelson, A.~Hoffman, E.~Osawa,
  et~al., {Growth, electronic properties and applications of nanodiamond},
  Diamond and Related Materials 17~(7-10) (2008) 1080--1088.
\newblock \href {https://doi.org/10.1016/j.diamond.2008.01.103}
  {\path{doi:10.1016/j.diamond.2008.01.103}}.

\bibitem{Mandal2019b}
S.~Mandal, H.~A. Bland, J.~A. Cuenca, M.~Snowball, O.~A. Williams,
  \href{http://xlink.rsc.org/?DOI=C9NR02729G}{{Superconducting boron doped
  nanocrystalline diamond on boron nitride ceramics}}, Nanoscale 11~(21) (2019)
  10266--10272.
\newblock \href {https://doi.org/10.1039/C9NR02729G}
  {\path{doi:10.1039/C9NR02729G}}.

\bibitem{Prawer2004a}
S.~Prawer, R.~J. Nemanich,
  \href{https://royalsocietypublishing.org/doi/10.1098/rsta.2004.1451}{{Raman
  spectroscopy of diamond and doped diamond}}, Philosophical Transactions of
  the Royal Society of London. Series A: Mathematical, Physical and Engineering
  Sciences 362~(1824) (2004) 2537--2565.
\newblock \href {https://doi.org/10.1098/rsta.2004.1451}
  {\path{doi:10.1098/rsta.2004.1451}}.

\bibitem{Sedov2019}
V.~S. Sedov, A.~K. Martyanov, A.~A. Khomich, S.~S. Savin, V.~V. Voronov, R.~A.
  Khmelnitskiy, et~al., \href{.}{{Co-deposition of diamond and $\beta$-SiC by
  microwave plasma CVD in H2-CH4-SiH4 gas mixtures}}, Diamond and Related
  Materials 98~(July) (2019) 107520.
\newblock \href {https://doi.org/10.1016/j.diamond.2019.107520}
  {\path{doi:10.1016/j.diamond.2019.107520}}.

\bibitem{Cuenca2022}
J.~A. Cuenca, S.~Mandal, E.~L.~H. Thomas, O.~A. Williams,
  \href{http://arxiv.org/abs/2111.10258}{{Microwave plasma modelling in
  clamshell chemical vapour deposition diamond reactors}}, Diamond and Related
  Materials (2022) Accepted\href {http://arxiv.org/abs/2111.10258}
  {\path{arXiv:2111.10258}}.

\bibitem{Mortet2017b}
V.~Mortet, Z.~{Vl{\v{c}}kov{\'{a}} {\v{Z}}ivcov{\'{a}}}, A.~Taylor, O.~Frank,
  P.~Hub{\'{i}}k, D.~Tr{\'{e}}mouilles, et~al., {Insight into boron-doped
  diamond Raman spectra characteristic features}, Carbon 115 (2017) 279--284.
\newblock \href {https://doi.org/10.1016/j.carbon.2017.01.022}
  {\path{doi:10.1016/j.carbon.2017.01.022}}.

\bibitem{Zibrov2018}
I.~P. Zibrov, V.~P. Filonenko, {Heavily boron doped diamond powder: Synthesis
  and rietveld refinement}, Crystals 8~(7) (2018) 1--7.
\newblock \href {https://doi.org/10.3390/cryst8070297}
  {\path{doi:10.3390/cryst8070297}}.

\bibitem{Nagasaka2016}
H.~Nagasaka, Y.~Teranishi, Y.~Kondo, T.~Miyamoto, T.~Shimizu,
  \href{https://www.jstage.jst.go.jp/article/ejssnt/14/0/14{\_}53/{\_}article}{{Growth
  Rate and Electrochemical Properties of Boron-Doped Diamond Films Prepared by
  Hot-Filament Chemical Vapor Deposition Methods}}, e-Journal of Surface
  Science and Nanotechnology 14~(March) (2016) 53--58.
\newblock \href {https://doi.org/10.1380/ejssnt.2016.53}
  {\path{doi:10.1380/ejssnt.2016.53}}.

\bibitem{Dubrovinskaia2006}
N.~Dubrovinskaia, L.~Dubrovinsky, N.~Miyajima, F.~Langenhorst, W.~A. Crichton,
  H.~F. Braun,
  \href{http://www.degruyter.com/view/j/znb.2006.61.issue-12/znb-2006-1213/znb-2006-1213.xml}{{High-pressure
  / High-temperature Synthesis and Characterization of Boron-doped Diamond}},
  Zeitschrift f{\"{u}}r Naturforschung B 61~(12) (2006) 1561--1565.
\newblock \href {https://doi.org/10.1515/znb-2006-1213}
  {\path{doi:10.1515/znb-2006-1213}}.

\bibitem{Cuenca2015b}
J.~A. Cuenca, {Characterisation of Powders Using Microwave Cavity
  Perturbation}, Ph.D. thesis, Cardiff University (2015).

\bibitem{Prozorov2006}
R.~Prozorov, R.~W. Giannetta, {Magnetic penetration depth in unconventional
  superconductors}, Superconductor Science and Technology 19~(8) (2006).
\newblock \href {https://doi.org/10.1088/0953-2048/19/8/R01}
  {\path{doi:10.1088/0953-2048/19/8/R01}}.

\bibitem{Klemencic2017b}
G.~M. Klemencic, J.~M. Fellows, J.~M. Werrell, S.~Mandal, S.~R. Giblin, R.~A.
  Smith, et~al.,
  \href{https://link.aps.org/doi/10.1103/PhysRevMaterials.1.044801}{{Fluctuation
  spectroscopy as a probe of granular superconducting diamond films}}, Physical
  Review Materials 1~(4) (2017) 1--5.
\newblock \href {http://arxiv.org/abs/1706.05845} {\path{arXiv:1706.05845}},
  \href {https://doi.org/10.1103/PhysRevMaterials.1.044801}
  {\path{doi:10.1103/PhysRevMaterials.1.044801}}.

\bibitem{Manifold2021}
S.~A. Manifold, G.~Klemencic, E.~L.~H. Thomas, S.~Mandal, H.~Bland, S.~R.
  Giblin, et~al., \href{https://doi.org/10.1016/j.carbon.2021.02.079}{{Contact
  resistance of various metallisation schemes to superconducting boron doped
  diamond between 1.9 and 300 K}}, Carbon 179 (2021) 13--19.
\newblock \href {https://doi.org/10.1016/j.carbon.2021.02.079}
  {\path{doi:10.1016/j.carbon.2021.02.079}}.

\bibitem{Hylton1989}
T.~L. Hylton, M.~R. Beasley, {Effect of grain boundaries on magnetic field
  penetration in polycrystalline superconductors}, Physical Review B 39~(13)
  (1989) 9042--9048.
\newblock \href {https://doi.org/10.1103/PhysRevB.39.9042}
  {\path{doi:10.1103/PhysRevB.39.9042}}.

\bibitem{Ishizaka2007}
K.~Ishizaka, R.~Eguchi, S.~Tsuda, T.~Yokoya, A.~Chainani, T.~Kiss, et~al.,
  \href{https://link.aps.org/doi/10.1103/PhysRevLett.98.047003}{{Observation of
  a Superconducting Gap in Boron-Doped Diamond by Laser-Excited Photoemission
  Spectroscopy}}, Physical Review Letters 98~(4) (2007) 047003.
\newblock \href {https://doi.org/10.1103/PhysRevLett.98.047003}
  {\path{doi:10.1103/PhysRevLett.98.047003}}.

\bibitem{Sacepe2006}
B.~Sac{\'{e}}p{\'{e}}, C.~Chapelier, C.~Marcenat, J.~Ka{\v{c}}mar{\v{c}}ik,
  T.~Klein, M.~Bernard, et~al., {Tunneling spectroscopy and vortex imaging in
  boron-doped diamond}, Physical Review Letters 96~(9) (2006) 1--4.
\newblock \href {http://arxiv.org/abs/0510541} {\path{arXiv:0510541}}, \href
  {https://doi.org/10.1103/PhysRevLett.96.097006}
  {\path{doi:10.1103/PhysRevLett.96.097006}}.

\bibitem{Wang2007}
Y.~Wang, H.~T. Su, F.~Huang, M.~J. Lancaster, {Measurement of YBCO thin film
  surface resistance using coplanar line resonator techniques from 20 MHz to 20
  GHz}, IEEE Transactions on Applied Superconductivity 17~(2) (2007)
  3632--3639.
\newblock \href {https://doi.org/10.1109/TASC.2007.899369}
  {\path{doi:10.1109/TASC.2007.899369}}.

\bibitem{Taber1990}
R.~C. Taber, \href{http://aip.scitation.org/doi/10.1063/1.1141389}{{A parallel
  plate resonator technique for microwave loss measurements on
  superconductors}}, Review of Scientific Instruments 61~(8) (1990) 2200--2206.
\newblock \href {https://doi.org/10.1063/1.1141389}
  {\path{doi:10.1063/1.1141389}}.

\bibitem{Kumar2017a}
D.~Kumar, M.~Chandran, M.~S. {Ramachandra Rao},
  \href{http://dx.doi.org/10.1063/1.4982591}{{Effect of boron doping on
  first-order Raman scattering in superconducting boron doped diamond films}},
  Applied Physics Letters 110~(19) (2017).
\newblock \href {https://doi.org/10.1063/1.4982591}
  {\path{doi:10.1063/1.4982591}}.

\bibitem{Zhang2019b}
W.~Zhang, K.~Kalashnikov, W.~S. Lu, P.~Kamenov, T.~Dinapoli, M.~E. Gershenson,
  {Microresonators Fabricated from High-Kinetic-Inductance Aluminum Films},
  Physical Review Applied 11~(1) (2019).
\newblock \href {http://arxiv.org/abs/1807.00210} {\path{arXiv:1807.00210}},
  \href {https://doi.org/10.1103/PhysRevApplied.11.011003}
  {\path{doi:10.1103/PhysRevApplied.11.011003}}.

\bibitem{Kalacheva2020}
D.~Kalacheva, G.~Fedorov, A.~Kulakova, J.~Zotova, E.~Korostylev, I.~Khrapach,
  et~al., {Improving the quality factor of superconducting resonators by
  post-process surface treatment}, AIP Conference Proceedings 2241~(June)
  (2020).
\newblock \href {https://doi.org/10.1063/5.0011900}
  {\path{doi:10.1063/5.0011900}}.

\bibitem{Hornibrook2012}
J.~M. Hornibrook, E.~E. Mitchell, C.~J. Lewis, D.~J. Reilly,
  \href{http://dx.doi.org/10.1016/j.phpro.2012.06.069}{{Parasitic losses in Nb
  superconducting resonators}}, Physics Procedia 36 (2012) 187--192.
\newblock \href {https://doi.org/10.1016/j.phpro.2012.06.069}
  {\path{doi:10.1016/j.phpro.2012.06.069}}.

\bibitem{Shao2003}
L.~Shao, J.~Liu, Q.~Y. Chen, W.~K. Chu, {Boron diffusion in silicon: The
  anomalies and control by point defect engineering}, Vol.~42, 2003.
\newblock \href {https://doi.org/10.1016/j.mser.2003.08.002}
  {\path{doi:10.1016/j.mser.2003.08.002}}.

\bibitem{Mirabella2013}
S.~Mirabella, D.~{De Salvador}, E.~Napolitani, E.~Bruno, F.~Priolo, {Mechanisms
  of boron diffusion in silicon and germanium}, Journal of Applied Physics
  113~(3) (2013).
\newblock \href {https://doi.org/10.1063/1.4763353}
  {\path{doi:10.1063/1.4763353}}.

\bibitem{Shenai1992}
K.~Shenai, {Diffusion Profiles of Boron Implanted into Plasma-Etched Silicon
  Surfaces}, IEEE Transactions on Electron Devices 39~(5) (1992) 1242--1245.
\newblock \href {https://doi.org/10.1109/16.129114}
  {\path{doi:10.1109/16.129114}}.

\end{thebibliography}

\end{document}